\documentclass[aps,prc,letterpaper,11pt,twoside,tightenlines,nofootinbib,showpacs,preprint]{revtex4-1}
\usepackage{graphicx}
\usepackage{epsfig,float}
\usepackage[sort&compress]{natbib}
\usepackage{subfigure}
\usepackage{amsmath}
\usepackage{amsfonts}
\usepackage{cancel}
\usepackage{lmodern,dsfont}
\usepackage{amssymb}
\usepackage{hyperref}
\begin{document}
\arraycolsep1.5pt
\newcommand{\Ima}{\textrm{Im}}
\newcommand{\Rea}{\textrm{Re}}
\newcommand{\mev}{\textrm{ MeV}}
\newcommand{\gev}{\textrm{ GeV}}
\newcommand{\dtres}{d^{\hspace{0.1mm} 3}\hspace{-0.5mm}}
\newcommand{\rts}{ \sqrt s}
\newcommand{\non}{\nonumber \\[2mm]}
\newcommand{\eps}{\epsilon}
\newcommand{\half}{\frac{1}{2}}
\newcommand{\thalf}{\textstyle \frac{1}{2}}
\newcommand{\Nmass}{M_{N}} 
\newcommand{\delmass}{M_{\Delta}} 
\newcommand{\pimass}{\mu}  
\newcommand{\rhomass}{m_\rho} 
\newcommand{\piNN}{f}      
\newcommand{\rhocoup}{g_\rho} 
\newcommand{\fpi}{f_\pi} 
\newcommand{\f}{f} 
\newcommand{\nucfld}{\psi_N} 
\newcommand{\delfld}{\psi_\Delta} 
\newcommand{\fpiNN}{f_{\pi N N}} 
\newcommand{\fpiND}{f_{\pi N \Delta}} 
\newcommand{\GMquark}{G^M_{(q)}} 
\newcommand{\vecpi}{\vec \pi}
\newcommand{\vectau}{\vec \tau}
\newcommand{\vecrho}{\vec \rho}
\newcommand{\delmu}{\partial_\mu}
\newcommand{\delMu}{\partial^\mu}
\newcommand{\nn}{\nonumber}
\newcommand{\bi}{\bibitem}
\newcommand{\vs}{\vspace{-0.20cm}}
\newcommand{\be}{\begin{equation}}
\newcommand{\ee}{\end{equation}}
\newcommand{\ba}{\begin{eqnarray}}
\newcommand{\ea}{\end{eqnarray}}
\newcommand{\ropi}{$\rho \rightarrow \pi^{0} \pi^{0}
\gamma$ }
\newcommand{\roeta}{$\rho \rightarrow \pi^{0} \eta
\gamma$ }
\newcommand{\omepi}{$\omega \rightarrow \pi^{0} \pi^{0}
\gamma$ }
\newcommand{\omeeta}{$\omega \rightarrow \pi^{0} \eta
\gamma$ }
\newcommand{\ul}{\underline}
\newcommand{\del}{\partial}
\newcommand{\rth}{\frac{1}{\sqrt{3}}}
\newcommand{\rsix}{\frac{1}{\sqrt{6}}}
\newcommand{\sq}{\sqrt}
\newcommand{\fr}{\frac}
\newcommand{\pr}{^\prime}
\newcommand{\ov}{\overline}
\newcommand{\Gm}{\Gamma}
\newcommand{\rw}{\rightarrow}
\newcommand{\rgl}{\rangle}
\newcommand{\De}{\Delta}
\newcommand{\Dp}{\Delta^+}
\newcommand{\Dm}{\Delta^-}
\newcommand{\Dz}{\Delta^0}
\newcommand{\Dpp}{\Delta^{++}}
\newcommand{\Sg}{\Sigma^*}
\newcommand{\Sp}{\Sigma^{*+}}
\newcommand{\Sm}{\Sigma^{*-}}
\newcommand{\Sz}{\Sigma^{*0}}
\newcommand{\X}{\Xi^*}
\newcommand{\Xm}{\Xi^{*-}}
\newcommand{\Xz}{\Xi^{*0}}
\newcommand{\Om}{\Omega}
\newcommand{\Omm}{\Omega^-}
\newcommand{\kp}{K^+}
\newcommand{\kz}{K^0}
\newcommand{\pip}{\pi^+}
\newcommand{\pim}{\pi^-}
\newcommand{\piz}{\pi^0}
\newcommand{\et}{\eta}
\newcommand{\kb}{\ov K}
\newcommand{\km}{K^-}
\newcommand{\kbz}{\ov K^0}
\newcommand{\ksb}{\ov {K^*}}

\def\tstrut{\vrule height2.5ex depth0pt width0pt} 
\def\jtstrut{\vrule height5ex depth0pt width0pt} 

\title{The role of $f_0(1710)$ in the $\phi \omega$ threshold peak of $J/\Psi \to \gamma \phi \omega$}

\author{A. Mart\'inez Torres$^{1}$,  K. P. Khemchandani$^{1}$, F. S. Navarra$^{1}$, M. Nielsen$^{1}$, and E. Oset$^{1,2}$ }
\affiliation{$^{1}$Instituto de F\'isica, Universidade de S\~ao Paulo, C.P. 66318, 05389-970 S\~ao 
Paulo, SP, Brazil.\\}
\affiliation{$^{2}$Departamento de F\'isica Te\'orica and IFIC, Centro Mixto Universidad de 
Valencia-CSIC,
Institutos de Investigaci\'on de Paterna, Aptdo. 22085, 46071 Valencia,
Spain\\}

\date{\today\\}

\begin{abstract}
We study the process $J/\Psi \to \gamma \phi \omega$, measured by the BES experiment, where a neat peak close to
the $\phi \omega$ threshold is observed and is associated to  a scalar meson resonance
around 1800 MeV. We make the observation that a scalar resonance coupling to $\phi
\omega$ unavoidably couples strongly to $K \bar K$, but no trace of a peak is seen in the
$K \bar K$ spectrum of the $J/\Psi \to \gamma K \bar K$ at this energy. This serves us to rule
out the interpretation of the observed peak as a signal of a new resonance. After this is
done, a thorough study is performed on the production of a pair of vector mesons and how its
interaction leads necessarily to a peak in the $J/\Psi \to \gamma \phi
\omega$ reaction close to the $\phi \omega$ threshold, due to the dynamical generation of
the $f_0(1710)$ resonance by the vector-vector interaction. We then show that both the
shape obtained for the $\phi \omega$ mass distribution, as well as the strength are
naturally reproduced by this mechanism. The work also explains why the $\phi\omega$ peak is observed in the BES experiment
and not  in other reactions, like $B^\pm\to K^\pm \phi\omega$ of Belle.

\end{abstract}

\maketitle

\section{Introduction}
\label{Intro}

A BES experiment looking for the decay of $J/\Psi \to \gamma \phi \omega$ \cite{besexp} observed a neat peak close to the $\phi \omega$ threshold which was tentatively associated to a $J^{PC}=0^{++}$ state with mass around 1812 MeV and width of about 105 MeV. An experiment with ten times more statistics has been reported recently \cite{expnew} and the peak is reconfirmed, the reanalysis leading to claims of a state with mass $M= 1795 \pm7^{+23}_{-5}$ MeV and a width $\Gamma = 95\pm10^{+78}_{-34}$ MeV, where the first error is statistical and the second systematic. They also report a product of branching ratios 
$B(J/\Psi \to \gamma R)~\times~B(R \to \phi\omega ) = (2.00\pm 0.08^{+1.38}_{-1.00})\times 10^{-4}$. The decay of $J/\Psi \to \gamma \phi \omega$ is doubly OZI suppressed with a production rate that is
smaller by at least one order of magnitude with respect to  $J/\Psi \to \gamma \omega \omega$ and $J/\Psi \to \gamma \phi \phi$ \cite{ozi}.

As usual, any new claim of a state is followed by theoretical suggestions for its interpretation, and in this case there have been works offering possible interpretations as a tetraquark state \cite{Li:2006ru}, a hybrid \cite{Chao:2006fq}, a glueball state \cite{Bicudo:2006sd}, a
threshold cusp attracting a resonance \cite{Bugg:2006uk} and an effect due to intermediate meson rescattering \cite{ihepth}. So far, none of these interpretations has been ruled out or supported by the experiment. 

In Ref.~\cite{Bicudo:2006sd} the $J/\Psi\to \gamma G$ is considered, where $G$ is a glueball state, followed by the decay of $G$ into vector-vector, which is studied in Refs.~\cite{cot2,cot3} within the vector meson dominance hypothesis. The possibility that the peak seen at BES could be a glueball is found likely, although other alternative explanations are not excluded.

In the present work we propose a different interpretation as due to the production of the  $f_0(1710)$ resonance below the $\phi \omega$ threshold. We shall show that the presence of this resonance necessarily leads to a peak around the  $\phi \omega$ threshold with a shape and strength compatible with experiment. Invoking the principle that if one phenomenon can be explained by an already established mechanism one should not make claims of new physics, we will conclude that the observed peak is not a signal of a new resonance but a manifestation of the  $f_0(1710)$. One might try to cast doubts on the peak seen in BES III since a devoted search of the Belle collaboration in the $B^{\pm} \to K^{\pm}\phi \omega $ reaction \cite{Liu:2009kca} does not see the peak. However, we shall provide an explanation of why these two facts are not contradictory.

\begin{figure}[h]
\begin{center}
\includegraphics[scale=0.6]{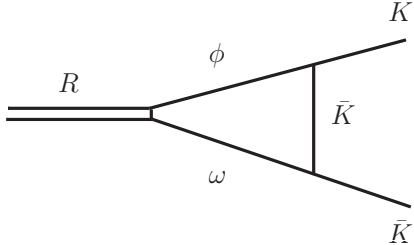}
\caption{A resonance $R$ of mass $\sim$ 1800 MeV, that couples to $\phi\omega$, decaying to $K\bar K$.}
\label{dec}
\end{center}
\end{figure}

 A very strong argument against the peak being a new resonance decaying to $\phi\omega$ is provided by the fact that both the $\omega$ and the  $\phi$ couple strongly to $K \bar K$. In this case the  $\omega$ and  $\phi$ can emit both a $K \bar K$ pair and one of the $K$ can be exchanged virtually between the two vectors, leading to a decay of the $0^{++}$ state into $K \bar K$ with $L=0$ (see Fig.~\ref{dec}). This is discussed in detail in section II. C and Fig. 2 of Ref.~\cite{gengvec}. There one can see that the largest part of the width of a resonance with mass around 1800 MeV coupling to $\phi\omega$ would be due to the decay into the $K\bar K$ channel  since there is practically no phase space for decay into $\phi\omega$ or $K^*\bar K^*$. In the $K\bar K$ decay channel the mass of the resonance would be very far from the $K\bar K$ threshold and the peak should be clearly observable, with no ambiguities about its interpretation\footnote{Actually, noting this fact, the authors of Ref.~\cite{Bicudo:2006sd} suggest that it would be interesting to look in detail for the $\gamma K\bar K$ decay channel. }. Yet, in the experiment studying $J/\Psi$ decay into $\gamma K \bar K$, clear peaks are seen for the $f_0(1500)$ and  $f_0(1710)$ but no trace is seen of any peak around 1800 MeV \cite{Bai:2003ww}. Similarly, MARK III~\cite{mark3} reports a clear signal for the $f_0(1710)$ in the $K\bar K$ spectra but no signal around 1800 MeV.

 At the same time a new resonance is being claimed with quantum numbers $0^{++}$ and mass close to 1800 MeV, observed in the two pion and four pion mass spectrum, named $f_0(1790)$~\cite{Bes1,Bes2,Bugg} and which one could be tempted to link to the peak found in $J/\Psi\to \gamma\phi\omega$.~This possibility can be, however, easily ruled out by considering the fact that the decay of $f_0(1790)$ to $K\bar K$ has been found to be strongly suppressed as compared to $\pi\pi$ or $\pi\pi\pi\pi$ channels (for a natural explanation of this property of the $f_0(1790)$ see Ref.~\cite{mkjh}).  Quoting textually from Ref.~\cite{Bes2}:  ``A particular feature is that the $f_0(1790)\to\pi\pi$ is strong, but there is little or no corresponding signal for decays to $K\bar K$". As a consequence, the enhancement observed near threshold in $J/\Psi\to\gamma \phi \omega$ can not be related with the $f_0(1790)$ since the $\phi\omega$ system would not have a suppressed decay to $K \bar K$ (see Fig.~\ref{dec}), but just the opposite, as explained a few lines above.

Among the theoretical papers mentioned, Ref.~\cite{ihepth} deserves a special attention since the idea is also that the peak observed could be a manifestation of the $f_0(1710)$ resonance, as we state here, a possibility also hinted in Ref.~\cite{Bicudo:2006sd}. The idea in Ref.~\cite{ihepth} is that the $J/\Psi$ decays into $\gamma f_0(1710)$, then the $f_0(1710)$ resonance couples to a pair of mesons (vector mesons, particularly $K^* \bar K^*$, were shown to be dominant) and there is rescattering of these vector mesons to produce the $\phi \omega$ final state. An enhancement close to the $\phi \omega$ threshold was produced with this mechanism, with a strength much smaller than the experimental data, and no firm conclusions were drawn. Yet, as the authors mentioned, there were many unknowns in the model, particularly tied to the interaction of vector mesons ($V$), where a perturbative approach was followed, assuming $VV \to \phi \omega$ transition mediated by $K$ and $\kappa$ exchange. An important step forward in this direction was taken later on in Refs.~\cite{raquel,gengvec}, where a thorough study of the vector-vector interaction was done using a coupled channel unitary approach with the dynamics extracted from the local hidden gauge Lagrangians \cite{hidden1,hidden2,hidden4}. This study allowed to see that the vector-vector scattering matrices develop some poles as a consequence of the interaction, and resonances are generated. These resonant states qualify as molecular states of two vector mesons and are usually referred to as dynamically generated resonances. Among many of them, the $f_0(1710)$ is generated and couples strongly to $K^* {\bar K}^*$ and $\phi \omega$. In such a case, the mechanism for final $\phi \omega$ production proceeds with a primary production of $\gamma VV$ followed by rescattering of these vectors to produce $\phi \omega$ in the final state. As we shall see, the primary production of $\phi \omega$ is not allowed and only the $VV$ rescattering leads to the $\phi \omega$ in the final state. One could reinterpret the doubly OZI suppressed mechanism for $\phi\omega $ production in this way, the first suppression applying to the production of all VV pairs without charm. This particular feature actually works in favor of having a more neat resonant shape since the 
$\phi \omega$ comes only from rescattering of the vector mesons by means of an amplitude that incorporates the  $f_0(1710)$ resonance. Thus, a background from uncorrelated $\phi \omega$ production is essentially absent in the mechanism of production and the effects of the $f_0(1710)$ show up more clearly. Since the resonance is below the $\phi \omega$ threshold, it is a combination of the tail of this resonance and the increasing phase space for $\phi \omega$ production what produces the visible peak. An enhancement of the strength near threshold due to resonances below threshold is unavoidable and this is a well known effect. Sometimes this shows up only as a deviation from phase space, with no peak structure  \cite{Dzyuba:2008fi,Oset:2001ep,MartinezTorres:2009cw}, but, depending on the strength of the background, sometimes it can also show up as a clear peak. This was the case of the $e^+e^-\to J/\Psi D \bar D$ reaction, where one peak close to the $D \bar D$ threshold was observed and associated to a resonance in the Belle collaboration work of  Ref.~\cite{Abe:2007sya}. However, in Ref.~\cite{Gamermann:2007mu} it was shown that a better fit to the data occurred due to the presence of a scalar hidden charm state below the $D\bar D$ threshold, X(3700), predicted in Ref.~\cite{dani}. In the present case, the absence of a significant background for $\phi \omega$ production magnifies the resonance shape close to the $\phi \omega$ threshold, to the point that in Refs.~\cite{besexp,expnew} a strong case was made about the discovery of a new resonance. We shall argue here that this is not the case, showing that the peak comes as an unavoidable consequence of the coupling of $\phi \omega$ to the $f_0(1710)$ resonance.

\section{Formalism}

In Ref.~\cite{gengchinos} the study of the radiative decay modes of the $J/\Psi$ into a photon and one of the tensor mesons $f_2(1270)$, $f'_2(1525)$, as well as the scalar ones $f_0(1370)$ and $f_0(1710)$, was undertaken and a good agreement with ratios of branching ratios was obtained. We will follow closely this formalism since for our present study we need both the  radiative decay of the $J/\Psi$ into $f_0(1710)$, as well as the more concrete one of the  $J/\Psi \to \gamma \phi \omega$.

\begin{figure}[b]
\begin{center}
\includegraphics[scale=0.6]{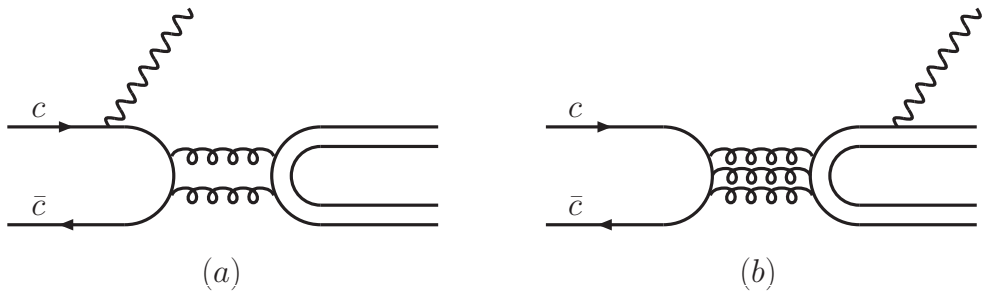}
\caption{Two mechanisms of the $J/\Psi$ radiative decay considering the possibilities that: (a) the photon is radiated from the initial $c\bar c$ state and (b) the photon is radiated from the final hadronic state.}
\label{f1}
\end{center}
\end{figure}

\begin{figure}[htpb]
\begin{center}
\includegraphics[scale=0.6]{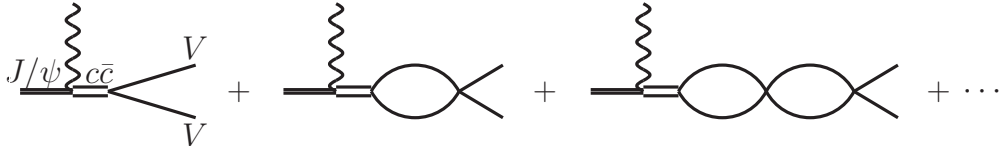}
\caption{Schematic representation of $J/\Psi$ decay into a photon and one dynamically generated resonance.} \label{f3}
\end{center}
\end{figure}

As in Ref.~\cite{gengchinos}, we assume, following the argumentation of Ref.~\cite{ozi}, that the mechanism of Fig.~\ref{f1}a
dominates the reaction. Further support for this assumption was found in Ref.~\cite{gengchinos}. Then, the $c\bar{c}$ component after the $\gamma$ radiation can decay into pairs of vectors, which inevitably will interact among themselves. This is shown diagrammatically in Fig.~\ref{f3}.
The following step is to recall that the $c\bar{c}$ object can be considered as an SU(3) singlet and then the pair of original vectors in the primary step will couple to an SU(3) singlet. The vector-vector content in the SU(3) singlet can be easily obtained from the trace of $ \mathrm{V\cdot V}$
\begin{equation}
 \mathrm{VV}_{\mbox{SU(3) singlet}}=\mathrm{Tr}[V\cdot V],
\end{equation}
where $V$ is the SU(3) matrix of the vector mesons
\begin{equation}
 V=\left(\begin{array}{ccc}
          \frac{1}{\sqrt{2}}\rho^0+\frac{1}{\sqrt{2}}\omega & \rho^+ &K^{*+}\\
           \rho^- &-\frac{1}{\sqrt{2}}\rho^0+\frac{1}{\sqrt{2}}\omega & K^{*0}\\
           K^{*-} &\bar{K}^{*0}&\phi
         \end{array}
\right).
\end{equation}
We, thus, find the vertex
\begin{equation}
 \mathrm{VV}_{\mbox{SU(3) singlet}}=
\rho^0\rho^0+\rho^+\rho^-+\rho^-\rho^++\omega\omega+K^{*+}K^{*-}+K^{*0}\bar{K}^{*0}
+K^{*-}K^{*+}+\bar{K}^{*0}K^{*0}+\phi\phi.
\end{equation}
One then projects this combination of VV  states, which are the
building blocks of the resonance produced, into VV isospin states with unitary normalization (an extra
factor $1/\sqrt{2}$ for identical particles or symmetrized ones) and
phase convention $|\rho^+\rangle=-|1,+1\rangle$, $|K^{*-}\rangle=-|1/2,-1/2\rangle$,
 \begin{eqnarray}
 |\rho\rho\rangle_\mathrm{I=0}&=&-\frac{1}{\sqrt{6}}|\rho^0\rho^0+\rho^+\rho^-+\rho^-\rho^+\rangle,\\
|K^*\bar{K}^*\rangle_\mathrm{I=0}&=&-\frac{1}{2\sqrt{2}}|K^{*+}K^{*-}+K^{*0}\bar{K}^{*0}+K^{*-}K^{*+}+\bar{K}^{*0}K^{*0}\rangle,\\
|\omega\omega\rangle_\mathrm{I=0}&=&\frac{1}{\sqrt{2}}|\omega\omega\rangle,\\
|\phi\phi\rangle_\mathrm{I=0}&=&\frac{1}{\sqrt{2}}|\phi\phi\rangle,
\end{eqnarray}
and one gets the weights for primary VV production of the process
$J/\Psi\rightarrow\gamma\mathrm{VV}$ with VV pairs in $I=0$:
\begin{equation}\label{eq:cg}
 w_i=\left\{\begin{array}{ll}
             -\sqrt{\frac{3}{2}}&\quad\mbox{for $\rho\rho$}\\
             -\sqrt{2}&\quad\mbox{for $K^*\bar{K}^*$}\\
             \frac{1}{\sqrt{2}} &\quad \mbox{for $\omega\omega$}\\
             \frac{1}{\sqrt{2}} &\quad\mbox{for $\phi\phi$}.
            \end{array}
\right. .
\end{equation}
It is interesting to note that there is no primary production of $\phi \omega$ with this mechanism. Production of $\phi \omega$ will occur with the rescattering of the primary $VV$ vectors as depicted in Fig.~\ref{f3}, and the sum of these terms is readily done by means of
\begin{equation}
t_{J/\Psi \rightarrow \gamma \phi \omega} = A \sum\limits_{j=1}^4 w_j G_j t_{j\rightarrow \phi \omega},\label{eq:tjpsi}
\end{equation}
with $A$ an unknown constant, $w_j$ the weights of Eq.~(\ref{eq:cg}) for the different primary $VV$ channels, $G_j$ the loop function for the intermediate $VV$ states and  $t_{j\rightarrow \phi \omega}$ the transition scattering matrix from the intermediate $VV$ states  to $\phi \omega$. We take the information for the $G_j$ and $t_{ij}$ functions from Ref.~\cite{gengvec}. 
 
 In Eq.~(\ref{eq:tjpsi})  $A$ represents the reduced matrix element for the operator responsible for the transition $c\bar c\to \textrm{VV}_{\textrm{SU(3) singlet}}$. Indeed
 
 \begin{align}
 \langle c\bar c| O_p|(VV)_j\rangle=\sum_R\langle c\bar c|O_p|(VV)_R\rangle\langle (VV)_R| (VV)_j\rangle,
 \end{align}
where $(VV)_R$ denotes a given $R$ representation of SU(3) for two vectors and $(VV)_j$ refers to a particular $VV$ physical channel. Since $c\bar c$ is an SU(3) singlet only the $\textrm{VV}_{\textrm{SU(3) singlet}}$ representation contributes to the sum. The factor $\langle c\bar c|O_p|(VV)_R\rangle$ stands for the coefficient $A$ and $\langle (VV)_R| (VV)_j\rangle$ are then the $\omega_j$ coefficients given in Eq.~(\ref{eq:cg}).

The argument about the absence of $\phi\omega$ tree level is quite powerful. Indeed, take now the Belle reaction mentioned in the introduction $B^\pm\to K^\pm \phi\omega$. This is a weak interaction process which does not conserve isospin, and, thus, SU(3). As a consequence we can not associate a unique SU(3) representation to the $\phi\omega$ system and then we can have tree level $\phi\omega$ background which would blur the appearance of a possible peak. We shall give an example of this at the end of the results section. In view of this argument, the facts that a peak in the $\phi\omega$ invariant mass is seen in the BES experiment  but not at Belle can be reconciled. Thus, the Belle finding cannot be used to cast doubts on the BES peak, which we do not question. Instead we show that it is unavoidable as a consequence of the presence of the $f_0(1710)$ resonance.

 The $t_{i\to j}$ matrices can be traced back to the results of Ref.~\cite{gengvec} by writing for each resonance 
\begin{equation}
t_{i\to j} = \frac{g_i g_j}{s- M^{2}_R + i M_R\Gamma_{R}} \label{eq:bw}
\end{equation}
where $g_i, g_j$ are the couplings of the resonance to the $i, j$ channels. We only need the $f_0 (1710)$ resonance here and the couplings are tabulated in Table~\ref{table:sum}. 
 \begin{table*}[h!]
      \renewcommand{\arraystretch}{1.5}
     \setlength{\tabcolsep}{0.3cm}
\caption{Couplings $g_k$'s appearing in Eq.~(\ref{eq:bw}), with $k$ one of the coupled channels:
$\rho\rho$, $K^*\bar{K}^*$, $\omega\omega$, $\phi\phi$, and $\phi\omega$. The units of these $g_k$'s are MeV.
\label{table:sum}}
\begin{center}
\begin{tabular}{cccccc}\hline\hline
$R$ & $\rho\rho$ & $K^*\bar{K}^*$ &$\omega\omega$ & $\phi\phi$& $\phi \omega$  \\\hline
$f_0(1710)$ &  $-1030 + i 1086$  & $7124+i96$  & $-1763 + i108$  & $-2493 -i204$ & $3010
  - i210$\\
 \hline\hline
    \end{tabular}
\end{center}
\end{table*}

With the amplitude of Eq.~(\ref{eq:tjpsi}), which depends on the invariant mass of $\phi \omega$, we can construct the $\phi \omega$ mass distribution given by
\begin{equation}
\frac{d\Gamma}{dM_{\rm inv}} = \frac{1}{(2\pi)^3} \frac{1}{ 4 M_{J/\Psi}^2} \,\,p_{\gamma} \bar{q}_{\omega} \mid t_{J/\Psi \rightarrow \gamma \phi \omega} \mid^2, \label{eq:totalgamma}
\end{equation}
where $p_{\gamma}$ and $\bar{q}_{\omega}$ are the photon momentum in the $J/\Psi$ rest frame and the $\omega$ momentum in the $\phi \omega$ rest frame, respectively
\begin{eqnarray}\nonumber
p_\gamma &=& \frac{\lambda^{1/2} \left( M_{J/\Psi}^2, 0, M_{\rm inv}^2 \right)}{2 M_{J/\Psi}};\\
\bar{q}_\omega &=& \frac{\lambda^{1/2} \left( M_{\rm inv}^2, m_\omega^2, m_\phi^2 \right)}{2 M_{\rm inv}}.\label{eq:momentums}
\end{eqnarray}

\begin{figure}[htbp]
\begin{center}
\includegraphics[scale=0.6]{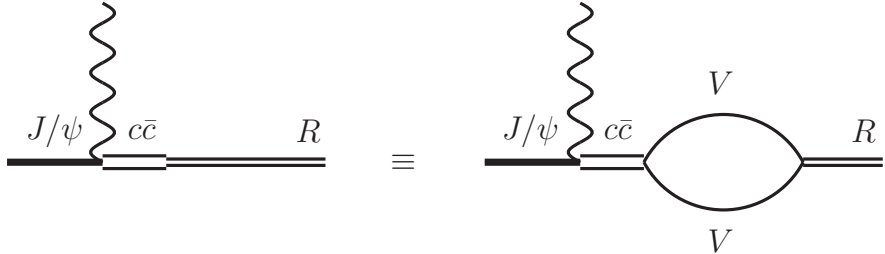}
\caption{Schematic representation of $J/\Psi$ decay into a photon and a dynamically generated resonance.} \label{f4}
\end{center}
\end{figure}

On the other hand, if we are interested in the production of the $f_0 (1710)$ resonance regardless of its decay channel, the relevant mechanism is depicted diagrammatically in Fig.~\ref{f4}
and we have 
\begin{equation}
t_{J/\Psi \rightarrow \gamma f_0 (1710)} = A \sum\limits_{j=1}^4 w_j G_j g_j \label{eq:tR}
\end{equation}
and the partial decay width for this process is given by
\begin{equation}
\Gamma_{J/\Psi  \rightarrow \gamma f_0(1710)} = \frac{1}{8\pi} \frac{1}{M_{J/\Psi}^2} \mid t_{J/\Psi \rightarrow \gamma f_0 (1710)} \mid^2 q_\gamma,\label{eq:f0_decayrate}
\end{equation}
where $q_\gamma$ is the momentum of the photon, like  $p_\gamma$ of Eq.~(\ref{eq:momentums}) but calculated at $M_{\rm inv} = M_{f_0(1710)}$. The theory cannot provide the value of the constant $A$ in Eqs.~(\ref{eq:tjpsi}) and (\ref{eq:tR}) since this requires a precise knowledge of the mechanism of OZI suppression, but if we divide $d\Gamma/ dM_{\rm inv}$ by $\Gamma_{J/\Psi \rightarrow \gamma f_0}$ the constant is cancelled and we can make precise predictions for the ratio. We shall call this ratio $R_{\Gamma}$
\begin{equation}
R_{\Gamma} = \frac{\displaystyle\int d M_{\rm inv} \dfrac{d\Gamma}{dM_{\rm inv}}}{\Gamma_{J/\Psi \rightarrow \gamma f_0 (1710)}}, \label{eq:ratio}
\end{equation}
which we can evaluate with the tools presented before. The value of this quantity is relevant because, together with the shape of the $\phi \omega$ mass distribution, it can be compared with the experimental values.

Experimentally we have, from the Particle Data Group (PDG) \cite{pdg},
\begin{equation}
B\left( J/\Psi \rightarrow \gamma f_0(1710) \rightarrow \gamma K \bar{K} \right) = \left( 8.5^{+1.2}_{-0.9} \right) \times 10^{-4}
\end{equation}
and for $B \left( f_0 (1710) \rightarrow K \bar{K} \right)$ we have some values in Ref.~\cite{pdg}
\begin{eqnarray}\nonumber
B \left( f_0 (1710) \rightarrow K \bar{K} \right) &= & 0.36 \pm 0.12 \hspace{1cm}\textrm {Ref.~\cite{alba}}\\\nonumber
&=& 0.38^{+0.09}_{-0.19}\hspace{1.5cm}\textrm {Ref.~\cite{longa}}\\
&=& 0.6 \hspace{2.5cm}\textrm {Ref.~\cite{ihepth}}.
\end{eqnarray}

Reference~\cite{alba} is a theoretical model where vector-vector and pseudoscalar-pseudoscalar channels are considered in a unitary way. Yet, the vector-vector interaction is omitted in the approach and it is this interaction what produces the $f_0(1710)$ resonance dynamically within the unitary treatment of the local hidden gauge approach~\cite{gengvec}, and it couples most strongly to $K^*\bar K^*$. Hence, the coupling to $K\bar K$ determined in Ref.~\cite{alba} must be used with caution. Reference~\cite{longa} is from 1986 and we assume that it has been improved by the number quoted in Ref.~\cite{ihepth} from 2006. We thus take the rate of Ref.~\cite{ihepth}, however, with an error to allow overlap with the numbers quoted in Refs.~\cite{alba} and \cite{longa}. Then we obtain
\begin{equation}
\frac{\Gamma_{J/\Psi \rightarrow \gamma f_0(1710)}}{\Gamma_{J/\Psi}} = (1.4^{+ 0.8}_{- 0.2})\times 10^{-3}.
\end{equation}
On the other hand, from Ref.~\cite{expnew} we have
\begin{equation}
B \left( J/\Psi \rightarrow \gamma R \rightarrow \gamma  \phi\omega \right) = \left( 2.00 \pm 0.08 ^{+ 1.38}_{-1.00} \right)\times 10^{-4},
\end{equation}
from where estimating roughly the errors we find
\begin{equation}
\frac{ B \left( J/\Psi \rightarrow \gamma R \rightarrow \gamma \phi \omega  \right)}{B \left( J/\Psi \rightarrow \gamma f_0 \right)} = 0.14^{+ 0.12}_{- 0.07}.\label{bb}
\end{equation}

\section{Results}
First of all we show the shape of the distribution $d \Gamma /d M_{\rm inv}$ and compare it with the updated data of the experiment \cite{expnew}. 

In our approach, and assuming the dominance of the diagram represented by Fig.~\ref{f1}a, there is no tree level contribution to the $\phi\omega$ production. However, in order to account for the strength of the distribution at large values of $M_{\rm inv}$, far away from the $f_0 (1710)$ resonance, we allow for a small background, which we take as a constant amplitude for simplicity, and we replace in Eq.~(\ref{eq:tjpsi})
\begin{equation}
t_{J/\Psi \rightarrow \gamma \phi \omega} \longrightarrow t_{J/\Psi \rightarrow \gamma \phi \omega}  + \beta, \label{eq:t+bkgnd}
\end{equation}
with $\beta$ being a constant (positive or negative) whose value is fixed by fitting the data around $M_{\rm inv} \simeq$ 3000 MeV where the $f_0 (1710)$ gives no relevant contribution.
The value of $\beta$ turns out to be of the same sign as Re$\{t_{J/\Psi \rightarrow \gamma \phi \omega}\}$.

\begin{figure}[htbp]
\begin{center}
\includegraphics[scale=0.6]{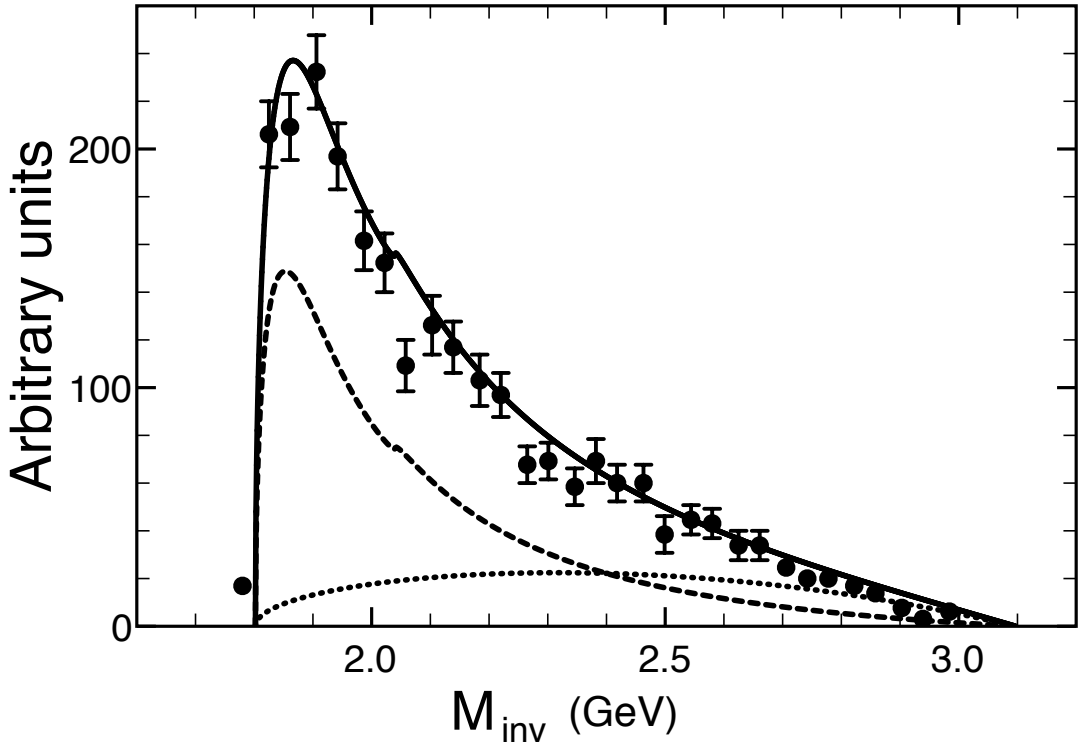}
\caption{The invariant mass distribution $\frac{d\Gamma}{d M_{inv}}$ for the process $J/\Psi \rightarrow \gamma \phi \omega$ from Eq.~(\ref{eq:totalgamma}). The data points, shown by filled circles, have been taken from Ref.~\cite{expnew}. The dotted and dashed lines represent the background  and the $f_0(1710)$ resonance contribution, respectively. The solid line shows the coherent sum of the two. } \label{invmass}
\end{center}
\end{figure}

In Fig.~\ref{invmass} we show the $\phi\omega$ invariant mass distribution obtained, fixing the total strength such as to reproduce
the peak of the experimental data on the number of $\phi \omega$ events per bin. As we can see, there is a perfect agreement between our results and the experimental data. This might be surprising at a first  sight, but the tail of the resonant shape of the amplitude of Eq.~(\ref{eq:tjpsi}), together with the phase space factors in Eq.~(\ref{eq:totalgamma}), essentially the factor $\bar{q}_\omega$ which vanishes at the $\phi \omega$ threshold, combine to give a peak close to the threshold. The resulting shape of the peak is linked to the dynamics of the process.

It is interesting to separate the contribution of the resonance and the background. In Fig.~\ref{invmass} we also show the contribution of the resonance alone, eliminating the background $\beta$ in Eq.~(\ref{eq:t+bkgnd}). As we can see, the resonance term is dominant, although the interference with the background raises the strength of the distribution. For comparison we also show in the figure what one obtains from the background alone.

\begin{figure}[htbp]
\begin{center}
\includegraphics[scale=0.6]{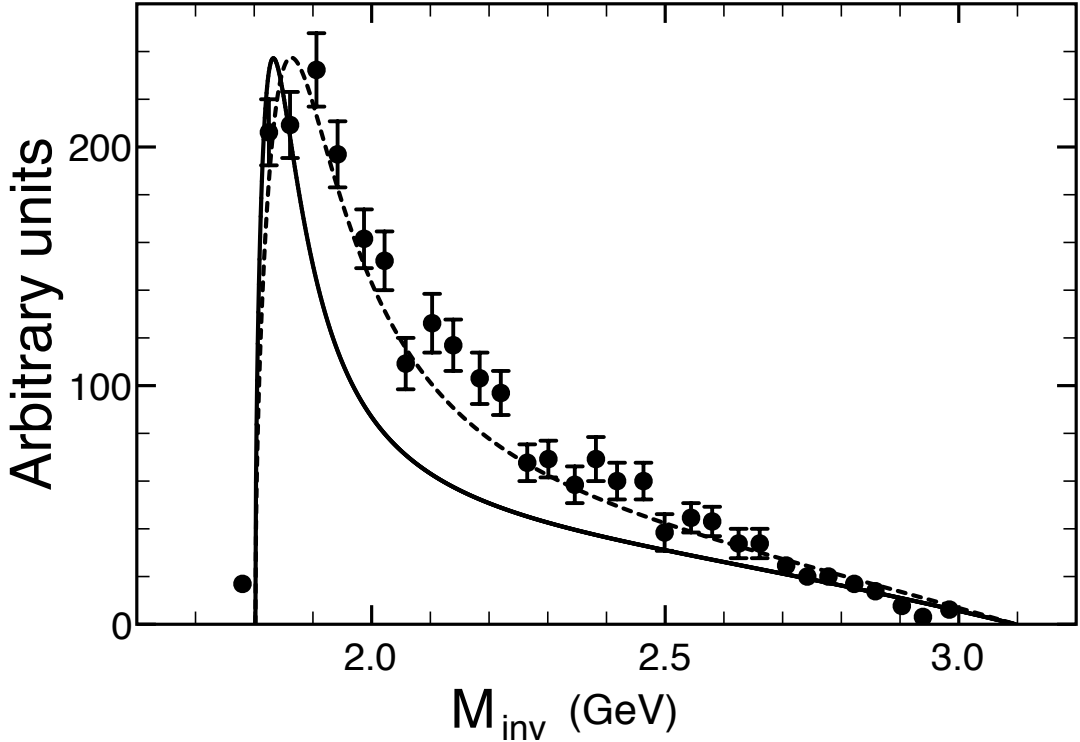}
\caption{The invariant mass distribution $\frac{d\Gamma}{d M_{inv}}$ for the empirical amplitude of Eq.~(\ref{empirical}). The solid line corresponds to the results found with a Breit-Wigner form with the central values of the resonance parameters suggested in Ref.~\cite{expnew}, i.e., mass of 1795 MeV and width of 95 MeV. The dashed curve corresponds to assuming the same mass but the upper limit of the width, 183 MeV.} \label{exppara}
\end{center}
\end{figure}

The agreement with the data is certainly a point to support the idea expressed in this paper. This agreement is better than the one that would be obtained by the resonance proposed in Ref.~\cite{expnew}. To show this, we simply substitute $t_{J/\Psi \rightarrow \gamma \phi \omega}$ of Eq.~(\ref{eq:tjpsi}) by 
\begin{equation}
t_{J/\Psi \to \gamma \phi \omega}  \to t_{\rm Emp} = \frac{A^{\prime}}{s- M^{2}_{R^\prime} + i  M_{R^\prime}
 \Gamma_{R^\prime}} + \beta^\prime\label{empirical}
\end{equation}
with $A^{\prime}$ adjusted to get the strength at the peak position and $\beta^\prime$ again adjusted to get the strength of the distribution at large values of $M_{\rm inv}$. After adjusting to the total strength of the peak we obtain the distribution shown in Fig.~\ref{exppara}, which does not reproduce well the data. 
It should be noted that in Ref.~\cite{expnew} a large range of mass and width of the resonance are given and we find that for the values of $\Gamma_{R^\prime}$ in the higher part of the range the agreement is better, yet of lower quality than the one provided by our approach. The important point is that once a new resonance is ruled out by the arguments in the introduction, the peak of  Ref.~\cite{expnew} is nicely reproduced in terms of the $f_0 (1710)$ dynamically generated in the approach of Ref.~\cite{gengvec}, independently that good fits could be obtained assuming a new resonance.

The work would not be complete if we did not calculate the strength of the $J/\Psi\to\gamma\phi\omega$ distribution. We do it now by evaluating the ratio of Eq.~(\ref{eq:ratio})
for the integrated $d\Gamma/dM_{\rm inv}$ and $\Gamma_{J/\Psi}\to\gamma f_{0}(1710)$. If we integrate up to $M^{\textrm{max}}_{\rm inv}=2.1$ GeV, a region that accounts for the largest part of the peak, we obtain a value of $R_{\Gamma}=0.15$, in agreement with the central experimental value of Eq.~(\ref{bb}). We can also quantify the theoretical uncertainties. First, we take the limit of integration to $M^{\textrm{max}}_{\rm inv}=2.3$ GeV, certainly an upper limit, and we get $R_{\Gamma}=0.21$. Further theoretical uncertainties can be obtained by changing randomly the mass and width of the $f_{0}(1710)$ between the range of the PDG, $1720\pm 6$ MeV and $135 \pm 8$ MeV, respectively, and the couplings $g_{i}$ by 10 $\%$. The ratio of 0.15 gets then converted into $R_{\Gamma}=0.15\pm0.04$, which added to the uncertainties in the choice of $M^{\textrm{max}}_{\rm inv}$ can be
set into $R_{\Gamma}=0.15^{+ 0.07}_{- 0.04}$.

As we can see, the range of theoretical values fully overlaps with the experimental one of Eq.~(\ref{bb}). This agreement is totally tied to the dynamics of the $VV$ interaction that we have used, and, in as much as this dynamics has been tested in so many processes \cite{gengchinos,albervv,junkogamgam,branzgamgam}, it stands on solid ground. Then, the agreement on the absolute values of the rate of production relative to $\Gamma_{J/\Psi}\to \gamma f_{0(1710)}$ is a strong point in favor of the idea exposed here that the peak observed in Refs.~\cite{besexp} and \cite{expnew} is due to the excitation of the $f_0 (1710)$ resonance and its further decay into $ \phi\omega$.

We have stressed the relevance of not having the $\phi\omega$ primary production to produce the shape of the experimental distribution. In order to further understand this point we have evaluated $\frac{d\Gamma}{d M_{inv}}$ for $K^{*}\bar K^{*}$ production where one has now a tree level contribution. The second term of Eq.~(\ref{eq:tjpsi}) would now be substituted by $A(\omega_{K^{*}\bar K^{*}}+\sum_{j=1}^{4} \omega_{j} G_{j} t_{j\to K^{*}\bar K^{*}})$ and $\bar q_{\omega}$ in Eq.~(\ref{eq:totalgamma}) by $\bar q_{K^{*}}$. In this case we observe that the background of the tree level largely dominates the distribution and only a very small peak at threshold appears that could be missed in an experiment with low resolution.

\section{Conclusions}
In this work we had a look at the data of two BES experiments for the reaction 
$J/\Psi \to \gamma \phi \omega$, where a neat peak is observed in the  $\phi \omega$ mass distribution close to the $\phi \omega$ threshold.  In the experimental works this peak was seen as a signal of one new scalar meson state with mass around 1800 MeV, not reported in the PDG. We made the important observation that both  $\phi$ and  $\omega$ couple strongly to $K \bar K$ (with the same strength) and a scalar resonance coupling to $\phi \omega$ unavoidably would couple to $K \bar K$, and it should be seen cleanly in the $K \bar K$ spectrum. The fact that no trace of a peak was seen in the experiment around this energy in the  $J/\Psi \to \gamma K \bar K$ reaction was proof enough to rule out the peak observed in $J/\Psi \to \gamma \phi\omega$ as a signal for a new state. We also noted that the lack of the tree level contribution for $\phi\omega$ production in the $J/\Psi \to \gamma \phi \omega$ reaction allowed to obtain a clear peak in the $\phi\omega$ invariant mass distribution. In this sense, we showed that in a reaction like in the Belle experiment $B^\pm\to K^\pm \phi\omega$~\cite{Liu:2009kca} there can be such $\phi\omega$ tree level contribution and the large background can dilute the peak, otherwise observed in the BES experiment.

  The main part of the work has then been devoted to show that the peak observed in the experiment is naturally obtained from the excitation of the $f_0(1710)$ resonance and its coupling to $\phi \omega$. The agreement of the $\phi \omega$ distribution with experiment was excellent and the absolute rates for the partial decay width of $J/\Psi \to \gamma \phi \omega$ reaction were also in very good agreement with experiment. The combination of all these facts clarifies the situation around this experiment with the conclusion that the peak observed is not a signal of a new resonance, but just a manifestation of the well established $f_0(1710)$ state.

\section*{Acknowledgments} 
The authors would like to thank Prof. David Bugg for useful discussions. 
This work is partly supported by the Spanish Ministerio de Economia y Competividad and European FEDER fund under the contract number
FIS2011-28853-C02-01 and the Generalitat Valenciana in the program Prometeo, 2009/090. We acknowledge the support of the European Community-Research Infrastructure
Integrating Activity
Study of Strongly Interacting Matter (acronym HadronPhysics3, Grant Agreement
n. 283286)
under the Seventh Framework Programme of EU. The authors would like to thank the Brazilian funding agencies FAPESP and CNPq for the financial support.

\end{document}